\documentclass[aps,pra,twocolumn,showpacs]{revtex4}

\usepackage{graphicx}
\usepackage{amsmath}
\usepackage{amssymb}
\usepackage{hyperref}

\begin{document}

  \title{Displacement-enhanced continuous-variable entanglement concentration}

  \author{Ond\v{r}ej \v{C}ernot\'ik}
  \author{Jarom\'ir Fiur\'a\v{s}ek}

  \affiliation{Department of Optics, Palack\'y University, 17. listopadu 12, 77146 Olomouc, Czech Republic}

  \date{\today}

  \begin{abstract}
    We study entanglement concentration of continuous variable Gaussian states by local photon subtractions enhanced by coherent displacements.
    Instead of the previously considered symmetric two-mode squeezed vacuum states, we investigate the protocol 
    for input states in the form of split single-mode squeezed vacuum,
    i.e., states obtained by mixing a single-mode squeezed vacuum with a vacuum state on a beam splitter,
    which is an experimentally highly relevant configuration.
    We analyze two scenarios in which the displacement-enhanced photon subtraction is performed either only on one, or on both of the modes
    and show that local displacements can lead to improved performance of the concentration protocol.
  \end{abstract}


  \maketitle

  \section{Introduction}

  In recent years, a great deal of effort has been dedicated to the study of continuous variable quantum information processing (CV QIP)
  that instead of qubits utilizes infinite dimensional Hilbert spaces of field modes.
  Of particular importance is the class of Gaussian states and Gaussian operations \cite{Braunstein05,Weedbrook12}
  which is experimentally feasible by linear optics, parametric amplifiers, and homodyne detection.
  Several important protocols such as quantum teleportation and quantum cryptography can be implemented using only Gaussian means
  \cite{Braunstein98,Ralph99,Silberhorn02} and have already been demonstrated experimentally \cite{Furusawa98,Grosshans03}.
  
  One of the key advantages of CV QIP is that the entangled quantum states \cite{Horodecki09},
  which represent a central ingredient of many quantum information processing tasks,
  can be generated deterministically using optical parametric amplifiers and beam splitters.
  However, entanglement distillation and concentration of Gaussian states belongs to the tasks 
  that require some non-Gaussian operation \cite{Eisert02,Fiurasek02,Giedke02}.
  Recall that entanglement distillation and concentration is a crucial tool 
  for entanglement enhancement and suppression of losses and noise in quantum communication. 
  This procedure enables two distant parties to extract, from a large number of shared weakly entangled  states,
  a smaller number of highly entangled states by means of local operations and classical communication \cite{Bennett95,Bennett96,Deutsch96}.
  Examples of non-Gaussian operations suitable for concentration of Gaussian entanglement include
  photon subtraction \cite{Opatrny00,Cochrane02,Olivares02,Datta12}, photon addition \cite{Parigi07,Navarrete-Benlloch12},
  interference with ancillary single photons \cite{Browne03,Eisert04,Lund09}, and Kerr nonlinearity \cite{Duan00,Fiurasek03}.

  Very recently, it was shown that, despite the fact that Gaussian operations cannot lead to concentration of Gaussian entanglement,
  they can improve performance of entanglement concentration based on photon subtraction \cite{Zhang11,Loock11,Fiurasek11}.
  Here, we follow this line of research but instead of a symmetric two-mode squeezed vacuum state,
  considered as an input state in  Refs.  \cite{Zhang11,Fiurasek11}, we analyze the protocol for a split single-mode squeezed vacuum state,
  i.e., single-mode squeezed vacuum mixed on a beam splitter with a vacuum state. 
  This latter configuration is highly relevant because the recent experimental demonstration of entanglement concentration by photon subtraction 
  was performed precisely with this kind of entangled state \cite{Takahashi10}.
  We assume a combination of local photon subtraction with coherent displacements, since squeezing, proposed in Ref. \cite{Zhang11},
  is experimentally much more challenging.
  Following the two entanglement-concentration schemes implemented by Takahashi \textit{et al.} \cite{Takahashi10}, we investigate configurations
  in which the photon subtraction is performed either by only one of the parties, or on both modes simultaneously.
  We show that while in the former scenario the displacements do not improve entanglement concentration,
  in the latter case the output entanglement is enhanced by the use of local displacements.

  The paper is organized as follows:
  We introduce the entanglement concentration protocol and the considered input states in Section~\ref{sec.protocol}.
  The main results are presented in Sections~\ref{sec.weak} and \ref{sec.arbitrary}, where the performance of the protocol is studied,
  first for weakly squeezed input states and later for states with arbitrary squeezing.
  Optimal displacements in dependence on the initial squeezing are determined,
  together with the resulting entanglement and success probability of the protocol.
  Finally, we conclude in Section~\ref{sec.conclusions}.
  
  \section{Entanglement concentration protocol}\label{sec.protocol}

\begin{figure}[t]
    \includegraphics[width=0.9\linewidth]{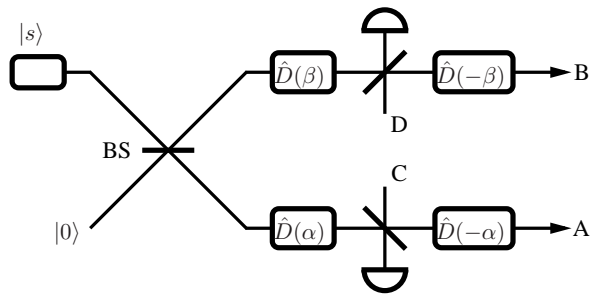}
    \caption{\label{fig.scheme}
      Entanglement concentration by photon subtraction enhanced by local displacements.
      The input state is obtained by mixing a single-mode squeezed vacuum state with vacuum on beam splitter BS.
      Prior to photon subtraction, local displacement $\hat{D}_A(\alpha)$ [$\hat{D}_B(\beta)$] is applied,
      which is undone after the subtraction by $\hat{D}_A^\dagger(\alpha)$ [$\hat{D}_B^\dagger(\beta)$].
      The subtraction is accomplished using a highly unbalanced beam splitter with an auxiliary mode C (D) in the vacuum state.
      A click of the on-off detector heralds a successful photon subtraction.
      The photon subtraction can be performed either only on one, or on both of the modes A, B.
    }
  \end{figure}

  The considered entanglement-concentration scheme is illustrated in Fig.~\ref{fig.scheme}.
  The input entangled Gaussian state is generated by splitting a single-mode squeezed vacuum (SMSV) state into two modes on a beam splitter BS.
  To simplify the subsequent analysis and obtain maximum insight into the studied protocol,
  we consider a pure input squeezed vacuum state with Fock-state expansion,
  \begin{equation}\label{eq.SMSV}
    |s\rangle = \frac{1}{\sqrt{\cosh s}}\sum_{n=0}^\infty \frac{\sqrt{(2n)!}}{2^n n!}(\tanh s)^n|2n\rangle,
  \end{equation}
  where $s$ denotes the squeezing constant.
  After mixing with vacuum on beam splitter  BS  with amplitude transmittance $t$ and reflectance $r$, we obtain the two-mode state
  \begin{equation}\label{eq.input}
    |\psi_\textrm{in}\rangle = \root 4\of{1-\lambda^2} \sum_{n=0}^\infty \sum_{k=0}^{2n} 
      \frac{\lambda^n}{2^n n!}\frac{(2n)!t^{2n-k}r^k}{\sqrt{k!(2n-k)!}} |2n-k,k\rangle,
  \end{equation}
  where $\lambda = \tanh s$ and $|m,n\rangle$ is a shorthand notation for $|m\rangle_A\otimes|n\rangle_B$.

  The two modes A and B are then distributed to two parties, Alice and Bob, who attempt to increase the entanglement of the shared state by local 
  coherent displacements $\hat{D}_A(\alpha)$ [$\hat{D}_B(\beta)$] and single photon subtractions.  
  As schematically indicated in Fig.~\ref{fig.scheme}, photon subtraction 
  is achieved using a highly unbalanced beam splitter which reflects a small part of the incoming  light onto a
  detector whose click heralds successful subtraction \cite{Ourjoumtsev06,Nielsen07,Sasaki07}. 
  Here and in Section~\ref{sec.weak}, we model photon subtraction by the action of an annihilation operator $\hat{a}$
  ($\hat{b}$) onto the state. A more realistic treatment taking into account reflectance of the tap-off beam splitters
  and properties of the detectors will be provided in Section~\ref{sec.arbitrary}.
  It will prove  helpful to assume that, after photon subtractions, Alice and Bob undo the displacements by applying inverse  displacements  
  $\hat{D}_A^\dagger(\alpha)=\hat{D}_A(-\alpha)$ [$ \hat{D}_B^\dagger(\beta)$] to the shared state.
  Note that these latter local unitary operations do not change the amount of entanglement so they are not a necessary part of 
  the entanglement concentration protocol and can be skipped in practice.

  In what follows, we shall consider two scenarios. 
  The first option is to perform the displacement-enhanced photon subtraction on one of the two modes only.
  In this case, the filtering operation reads
  \begin{equation}\label{eq.single}
    \hat{F}_1 = \hat{D}_A^\dagger(\alpha)\hat{a}\hat{D}_A(\alpha)\otimes\hat{\mathbb{I}}_B = (\hat{a}+\alpha)\otimes\hat{\mathbb{I}}_B,
  \end{equation}
  where $\hat{\mathbb{I}}$ denotes the identity operator.
  Secondly, we can subtract photons from both modes, which is experimentally more challenging, but brings another degree of freedom
  that can be used to optimize the protocol.
  In this situation, we get the filtering operator
  \begin{eqnarray}\label{eq.double}
    \hat{F}_2 &=& \hat{D}_A^\dagger(\alpha)\hat{a}\hat{D}_A(\alpha)\otimes\hat{D}_B^\dagger(\beta)\hat{b}\hat{D}_B(\beta) \nonumber \\
      &=& (\hat{a}+\alpha)\otimes(\hat{b}+\beta).
  \end{eqnarray}

  \section{Weak initial squeezing}\label{sec.weak}

  In this Section, we assume that the input state is only weakly squeezed, $\lambda \ll 1$, so that 
  it can be approximated as
  \begin{equation}\label{eq.weakSMSV}
    |\psi_{\mathrm{in}}\rangle \approx |0,0\rangle + \lambda rt|1,1\rangle + \frac{\lambda}{\sqrt{2}}(t^2|2,0\rangle+r^2|0,2\rangle).
  \end{equation}
  This truncated expansion contains the dominant vacuum term and a two-photon contribution originating from the two-photon component of $|s\rangle$
  split on the beam splitter BS.
  To quantify the entanglement of the filtered state $ |\psi_{AB}\rangle \propto \hat{F}|\psi_{\mathrm{in}}\rangle$, 
  we use the entropy of entanglement $E_S$ defined as the von Neumann entropy of one of the reduced states, 
  $E_S= S(\hat{\rho}_A)=S(\hat{\rho}_B)$,  where 
  $\hat{\rho}_{A}=\mathrm{Tr}_{B}[|\psi_{AB}\rangle\langle \psi_{AB}|]$,
  and $\hat{\rho}_B$ is defined similarly.   
  
  \begin{figure}[t]
    \includegraphics[width=0.8\linewidth]{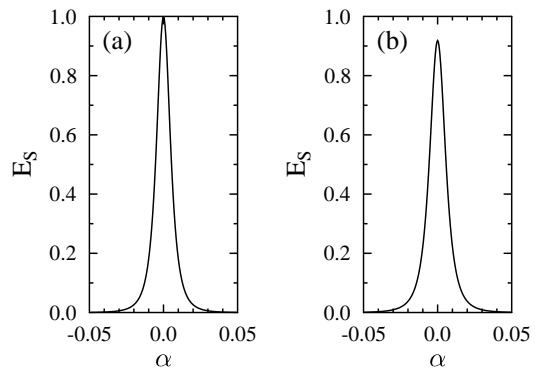}
    \caption{\label{fig.weak_one}
      Dependence of entropy of entanglement $E_S$ of transformed state on coherent displacement $\alpha$ with single-mode filtration.
      Input states were obtained by mixing the SMSV state with vacuum on a beam splitter with reflectance (a) $r = 1/\sqrt{2}$
      (b) and $r = 1/\sqrt{3}$.
      The input squeezing is $\lambda = 0.01$.
    }
  \end{figure}

  For the single-mode filtration (\ref{eq.single}), the resulting state $|\psi_1\rangle=\hat{F}_1|\psi_{\mathrm{in}}\rangle$ reads
  \begin{eqnarray}
    |\psi_1\rangle &\approx& \alpha|0,0\rangle + \lambda rt|0,1\rangle + \lambda t^2|1,0\rangle \nonumber \\
      &&+\, \alpha\lambda rt|1,1\rangle + \frac{\alpha\lambda}{\sqrt{2}}(t^2|2,0\rangle+r^2|0,2\rangle).
  \end{eqnarray}
  This formula suggests that the best choice of displacement is $\alpha = 0$, 
  together with using a balanced beam splitter for the initial state preparation,
  as this gives a maximally entangled two-qubit state $\frac{1}{\sqrt{2}}(|0,1\rangle+|1,0\rangle)$, i.e., a single-photon state 
  evenly split between modes A and B. We have carried out extensive numerical calculations that support this conjecture.
  As an example, obtained by numerical calculations truncated at the Fock number $n_\mathrm{max} = 10$,
  Fig.~\ref{fig.weak_one} shows the dependence of $E_S$ on displacement $\alpha$
  for two different beam splitter reflectances, namely, $r = 1/\sqrt{2}$ (left) and $r = 1/\sqrt{3}$ (right).
  Independent of the beam splitter reflectance, we get maximum entanglement for zero displacement.
  Furthermore, in this case the normalized filtered state is given by $ r|0,1\rangle + t|1,0\rangle$, 
  whose entanglement is maximized for $r = t = 1/\sqrt{2}$.
  
\begin{figure}[!b!]
    \includegraphics[width=0.8\linewidth]{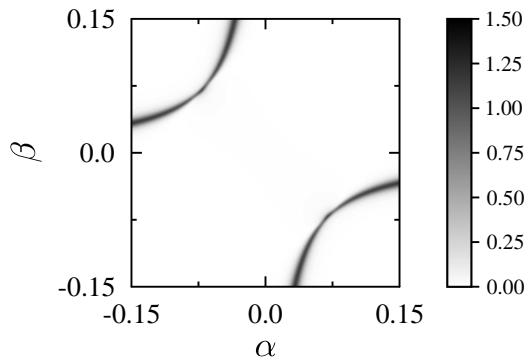}
    \caption{\label{fig.weak_two}
      The entropy of entanglement for the two-mode subtraction scheme.
      The output states with increased entanglement lie near a hyperbole $\beta = -\lambda rt/\alpha$.
      On the hyperbole, the dominant vacuum term vanishes completely, giving maximum output entanglement.
      The input squeezing is $\lambda = 0.01$ and the use of a balanced beam splitter, $r = 1/\sqrt{2}$, is assumed.
    }
  \end{figure}

  With photon subtraction performed on both modes, we get another degree of freedom, the second coherent displacement $\beta$,
  which can be used to maximize the entanglement of the output state. Using the input-state approximation (\ref{eq.weakSMSV}), the state
  $|\psi_2\rangle=\hat{F}_2|\psi_{\mathrm{in}}\rangle$ reads
  \begin{eqnarray}\label{eq.weak_two}
  |\psi_{2}\rangle&\approx&(\lambda rt + \alpha\beta)|0,0\rangle + \lambda(\alpha r + \beta t)(t|1,0\rangle + r|0,1\rangle) \nonumber \\
      &&+ \frac{\lambda}{\sqrt{2}}\alpha\beta(t^2|2,0\rangle +\sqrt{2}rt|1,1\rangle+ r^2|0,2\rangle).
      \label{psi2}
  \end{eqnarray}
  The entanglement of the output state is shown in Fig.~\ref{fig.weak_two}.
  We can see that the entanglement is maximized for displacements near the hyperbole $\alpha\beta = -\lambda rt$.
  If this condition is satisfied, the vacuum term in state $|\psi_2\rangle$ disappears due to destructive interference.
  For $\lambda \ll 1$ the state then effectively becomes the entangled single-photon state, 
  $|\psi_2\rangle \approx t|1,0\rangle + r|0,1\rangle$ because the last term becomes proportional to $\lambda^2$, hence negligible.
  This may look disappointing because  this  state can be obtained 
  much more easily by subtracting only a single photon.
   
\begin{figure}[!t!]
    \includegraphics[width=\linewidth]{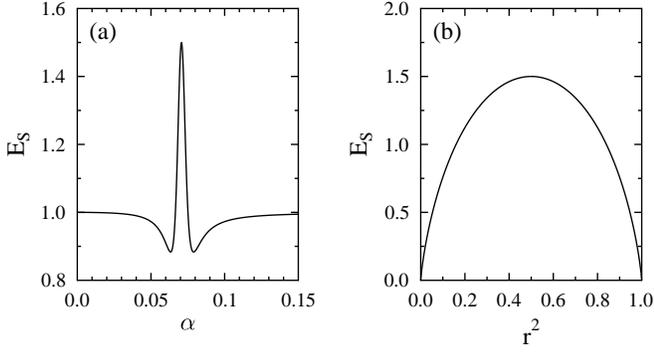}
    \caption{\label{fig.rec}
      Two-mode entanglement concentration with reciprocal displacements, $\beta = -\lambda rt/\alpha$.
      In panel (a), the dependence of the entropy of entanglement on displacement $\alpha$ is plotted for a balanced beam splitter.
      The plateaus with $E_S = 1$ correspond to the qubit state $t|1,0\rangle + r|0,1\rangle$,
      while a narrow peak appears for the qutrit state (\ref{eq.qutrit})
      when the one photon terms in (\ref{eq.weak_two}) vanish due to destructive interference.
      The entropy of entanglement of the qutrit state (\ref{eq.qutrit}) as a function of the beam splitter intensity reflectance $R = r^2$
      is represented in panel (b).
    }
  \end{figure}

  However, there is a further narrow peak clearly visible in Fig.~\ref{fig.rec}
  which emerges at $\alpha=\sqrt{\lambda}t$ and $\beta=-\sqrt{\lambda}r$. At this point, the single-photon part of $|\psi_2\rangle$
  is also eliminated by destructive interference as $\alpha r+\beta t=0$, so the two-photon part of $|\psi_2\rangle$ becomes dominant.
  Since this part is proportional to $\lambda^2$, we must take into account the four-photon terms 
  in the expansion of $|\psi_{\mathrm{in}}\rangle$ in our analysis. After photon subtractions, the dominant part of this term which is 
  proportional to $\lambda^2$ reads
  \begin{widetext}
    \begin{eqnarray}
      \hat{a}\hat{b}\sum_{k=0}^4  \frac{3\lambda^2t^{4-k}r^k}{\sqrt{(4-k)! k!}}|4-k,k\rangle 
        =\frac{3}{\sqrt{2}}\lambda^2 rt (t^2|2,0\rangle +\sqrt{2}rt|1,1\rangle+ r^2|0,2\rangle).
    \end{eqnarray}
  \end{widetext}
 Remarkably, this term has the same structure as the two-photon part of $|\psi_2\rangle$ in Eq. (\ref{psi2}).  
  Therefore, if $\alpha=\sqrt{\lambda}t$ and $\beta=-\sqrt{\lambda}r$  the state after filtration becomes an entangled two-photon state
  \begin{equation}\label{eq.qutrit}
    |\psi_2\rangle \approx \sqrt{2}rt|1,1\rangle + t^2|2,0\rangle + r^2|0,2\rangle.
  \end{equation} 
  The dependence of the entropy of entanglement of this state on the beam splitter reflectance is plotted in Fig.~\ref{fig.rec}(b).
  For a balanced beam splitter, we get $E_S = 1.5$, which is not far from the maximally entangled qutrit state,
  for which we have $E_S = \log_2 3 = 1.585$. We can conclude that the coherent displacements  improve the performance of 
  the scheme with two subtractions and allow us to generate more entanglement than the scheme with a single 
  subtraction only. The price for this entanglement gain is the reduced success rate that scales as $\lambda^2$
  for single subtraction while it becomes proportional to $\lambda^4$ for double subtraction. 
  The success probability and performance of the protocol
  are further analyzed in the next section where we extend our investigation beyond the weak-squeezing limit.

  \section{Entanglement concentration with arbitrarily squeezed states}\label{sec.arbitrary}

  We now provide a more generic analysis allowing for arbitrarily strong squeezing and accounting for the actual implementation 
  of photon subtraction as sketched in Fig.~\ref{fig.scheme}.
  We shall assume that the tap-off beam splitters have amplitude transmittance $t_S$ and reflectance $r_S$ with $t_S^2+r_S^2=1$. 
  The avalanche photodiodes commonly employed in this kind of experiments cannot distinguish the number of detected photons 
  and respond with a binary outcome (click or no click). 
  Consequently, the output state after photon subtraction becomes a mixture of states 
  corresponding to subtractions of different numbers of photons.
  Since our goal is to evaluate the entanglement of the output state, we continue the calculations in the Fock state representation. 
  We first derive the output state for single-mode photon subtraction; the extension to the two-mode subtraction is straightforward.

  It is convenient to rewrite the input two-mode state as
  \begin{equation}
    |\psi_{\mathrm{in}}\rangle=\sum_{m,n=0}^\infty C_{mn}|m,n\rangle,
  \end{equation}
  where
  \begin{equation}\label{eq.Cmatrix}
    C_{mn} = \left\{\begin{array}{ll} 
      \root 4\of{1-\lambda^2}\Big(\dfrac{\lambda}{2}\Big)^{\tfrac{m+n}{2}} \dfrac{(m+n)!}{(\frac{m+n}{2})!} \dfrac{t^m r^n}{\sqrt{m!n!}},
        & m+n\ \textrm{even}, \\
      0, & m+n\ \textrm{odd}.
    \end{array}\right.
  \end{equation}
  Let us now determine a pure state $|\psi^{(k)}\rangle$ of modes A and B after coherent displacement $\hat{D}_A(\alpha)$ 
  and conditional subtraction of $k$ photons from mode A. After some algebra one finds that
  \begin{equation}
    |\psi^{(k)}\rangle=\sum_{m,n=0}^\infty B_{mn}^{(k)}|m,n\rangle,
  \end{equation}
  where
  \begin{equation}\label{eq.subtracted}
    B_{mn}^{(k)} = \sqrt{{m+k}\choose k} t_S^m r_S^k \sum_{a=0}^\infty D_{m+k,a}(\alpha) C_{an},
  \end{equation}
   and 
  \begin{equation}
  D_{m,n}(\alpha)=e^{-\frac{|\alpha|^2}{2}}\sum_{k=k_0}^m\frac{\sqrt{m! \,n!\,}\alpha^k(-\alpha^\ast)^{n-m+k}}{k!\,(m-k)!\,(n-m+k)!}
  \end{equation}
  with $k_0=\max(0,m-n)$ is the matrix element of the displacement operator $\hat{D}(\alpha)$.
  If the detector which heralds successful photon subtraction only distinguishes between presence and absence of photons,
  the resulting state of modes A and B is given by a mixture of all states corresponding to subtraction of one or more photons,
  \begin{equation}
    \hat{\rho}_1 = \frac{1}{P_1} \sum_{k=1}^\infty |\psi^{(k)}\rangle\langle\psi^{(k)}|.
  \end{equation}
  The states $|\psi^{(k)}\rangle$ are not normalized and the norm of  $|\psi^{(k)}\rangle$ is equal to the probability of subtracting $k$ photons.
  The overall success probability therefore reads $P_1 = \sum_{k=1}^\infty \langle\psi^{(k)}|\psi^{(k)}\rangle$.

  A limited  efficiency $\eta$ of the heralding detectors can be modeled by a beam splitter with intensity transmittance 
  $\eta$ followed by  a perfect detector with unit efficiency.
  If $k$ photons are reflected on the tap-off beam splitter, then the inefficient detector does not register any photon 
  with probability $(1-\eta)^k$, and clicks with probability $1-(1-\eta)^k$.  The output state thus reads
  \begin{equation}\label{eq.onemode}
    \hat{\rho}_1(\eta) = \frac{1}{P_1(\eta)} \sum_{k=1}^\infty [1-(1-\eta)^k]|\psi^{(k)}\rangle\langle\psi^{(k)}|,
  \end{equation}
  where $P_1(\eta)= \sum_{k=1}^\infty [1-(1-\eta)^k]\langle\psi^{(k)}|\psi^{(k)}\rangle$.
  
  Going beyond transformation of specific input states,
  we next provide an explicit expression for a non-unitary quantum filter describing the combined effect 
  of coherent displacement and subtraction of exactly $k$ photons from the mode. 
  Similarly as in preceding sections, it is helpful to assume that the coherent displacement $\hat{D}(\alpha)$
  is undone after subtraction. Taking into account the reflectance of the tap-off beam splitter,
  the optimal inverse displacement reads $\hat{D}(-t_S\alpha)$.
  The formula for the filtering operator $\hat{F}^{(k)}$ is most easily derived by considering an input coherent state $|\gamma\rangle$. 
  We use the properties of  displacement operator $\hat{D}(\gamma)|0\rangle=|\gamma\rangle$ 
  and $\hat{D}(\alpha)\hat{D}(\gamma)=e^{(\alpha\gamma^\ast-\alpha^\ast\gamma)/2}\hat{D}(\alpha+\gamma)$ 
  and the transformation of the coherent state on the tap-off beam splitter,
  $|\xi\rangle _A|0\rangle_C\rightarrow |t_S\xi\rangle_A|r_S\xi\rangle_C$, 
  where C is the auxiliary vacuum input port of that beam splitter.  
  In this way we obtain 
  \begin{eqnarray}
    \hat{F}^{(k)}|\gamma\rangle&=& 
      e^{\frac{r_S^2}{2}(\alpha\gamma^\ast-\alpha^\ast\gamma)}{}_C\langle k|r_S(\alpha+\gamma)\rangle_C  |t_S\gamma\rangle_A \nonumber \\
      &=& e^{\frac{r_S^2}{2}(\alpha\gamma^\ast-\alpha^\ast\gamma)} e^{-\frac{r_S^2}{2}|\alpha+\gamma|^2}\frac{r_S^k}{\sqrt{k!}}(\alpha+\gamma)^k  
      |t_S\gamma\rangle \nonumber, 
  \end{eqnarray}
  which is valid for arbitrary $|\gamma\rangle$.
  From this expression we can determine the filtering operator,
  \begin{equation}
     \hat{F}_1^{(k)} = \frac{r_S^k}{\sqrt{k!}}e^{-\frac{r_S^2}{2}|\alpha|^2} t_S^{\hat{n}_A} e^{-r_S^2 \alpha^\ast \hat{a}}(\hat{a}+\alpha)^k.
  \end{equation}
  If we set $\alpha=0$ we recover filter describing $k$-photon subtraction \cite{Fiurasek09} 
  and in the limit $r_S\rightarrow 0$ we obtain the $k$th power of displaced annihilation operator $(\hat{a}+\alpha)^k$. 
  The factor $t_S^{\hat{n}}$ describes state attenuation on a beam splitter with transmittance $t_S$ 
  and an additional term $e^{-r_S^2 \alpha^\ast \hat{a}}$ emerges for nonzero displacement $\alpha$.

  The output state for the two-mode photon subtraction can be determined in a similar manner as for the single-mode subtraction. 
  After some algebra, we get
  \begin{eqnarray}\label{eq.twomode}
    \hat{\rho}_2 &=& \frac{1}{P_2(\eta)} \sum_{k,l=1}^\infty [1-(1-\eta)^k][1-(1-\eta)^l] \times \nonumber \\
      &&|\psi^{(k,l)}\rangle\langle\psi^{(k,l)}|, 
  \end{eqnarray}
  where state $|\psi^{(k,l)}\rangle = \sum_{m,n=0}^\infty B_{mn}^{(k,l)}|m,n\rangle$ is obtained
  after subtracting $k$ photons from mode A and $l$ photons from mode B,
  \begin{eqnarray}
    B_{mn}^{(k,l)} &=& \sqrt{{{m+k}\choose k}{{n+l}\choose l}} t_S^{m+n} r_S^{k+l} \times \nonumber \\
      && \sum_{a,b=0}^\infty D_{m+k,a}(\alpha)D_{n+l,b}(\beta) C_{ab},
  \end{eqnarray}
  and 
  \[
    P_2 = \sum_{k,l=1}^\infty [1-(1-\eta)^k][1-(1-\eta)^l]\langle\psi^{(k,l)}|\psi^{(k,l)}\rangle.
  \]
  Note that we assume the same reflectance $r_S$ of both tap-off beam splitters and equal detection efficiency $\eta$ of both detectors.

  To quantify the entanglement of the output mixed state, we use logarithmic negativity, defined as \cite{Vidal02}
  \begin{equation}
    E_N(\hat{\rho}) = \log_2 ||\hat{\rho}^{T_A}||_1,
  \end{equation}
  which is a computable entanglement measure for mixed states.
  Here, $T_A$ denotes partial transpose of density matrix $\hat{\rho}$ with respect to mode A and $||\cdot||_1$ denotes trace norm,
  $||\hat{X}||_1 = \textrm{Tr}\sqrt{\hat{X}^\dagger\hat{X}}$.

\begin{figure}
    \includegraphics[width=\linewidth]{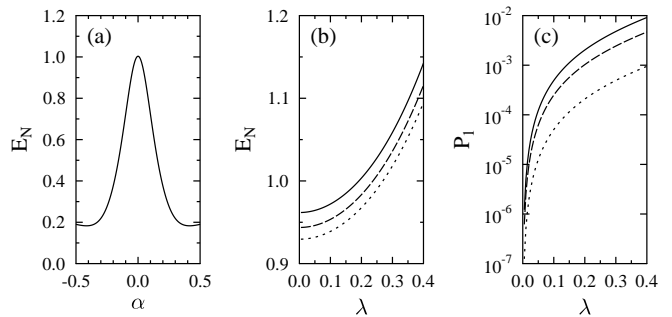}
    \caption{\label{fig.onemode}
      Performance of entanglement concentration for the single-mode subtraction scheme.
      Panel (a) shows logarithmic negativity as a function of displacement $\alpha$ for initial squeezing $\lambda = 0.2$.
      Zero displacement gives, as in the case of weak squeezing, the most entanglement on the output.
      Panels (b) and (c) show the logarithmic negativity and success probability, respectively,
      as functions of the initial squeezing $\lambda$ for zero displacement.
      The results were obtained for $R_S = r_S^2 = 0.1$,
      considered detection efficiencies are $\eta = 1$ (solid line), $\eta = 0.5$ (dashed) and $\eta = 0.1$ (dotted).
    }
  \end{figure}

  In numerical calculations, the infinite summations are truncated at sufficiently large Fock number $n_{\mathrm{max}}$,
  where $n_\mathrm{max} = 10$ is sufficient for the values of squeezing considered in this paper.
  Numerical results obtained for photon subtraction performed on one mode are plotted in Fig.~\ref{fig.onemode}.
  We consider balanced beam splitter for preparation of the input entangled state.
  In fact, the output entanglement is maximized for a slightly unbalanced beam splitter, as the single-mode filtration is asymmetric,
  but the change of reflectance is negligible (approximately 0.5 \% higher intensity reflectance for $\lambda = 0.2$ and $r_S^2 = 0.1$).
  As in the case of weakly squeezed input states, the most entanglement is achieved with zero displacement; see Fig.~\ref{fig.onemode}(a).
  In the weak-squeezing limit, $E_N$ is limited by the reflectance of the tap-off beam splitter and the detector efficiency 
  and more than 0.9 e-bit of entanglement can be extracted even for realistic parameters $r_S^2=0.1$ and $\eta=0.1$.
  As shown in Fig.~\ref{fig.onemode}(b), the maximum entanglement 
  grows only weakly with increasing squeezing ($\lambda=0.4$ corresponds to more than $3$~dB of squeezing). On the other hand,
  the amount of squeezing has a strong impact on the success probability of the photon subtraction. 
  In the considered setup with highly a unbalanced tap-off beam splitter, 
  the limited efficiency of the trigger detector mainly decreases the success probability that is approximately proportional to $\eta$,
  while the entanglement is not much reduced.

  \begin{figure}
    \includegraphics[width=\linewidth]{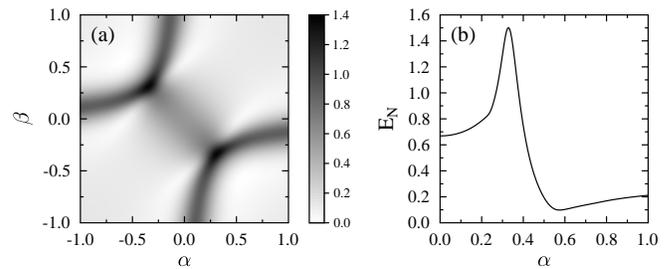}
    \caption{\label{fig.twomode}
      Entanglement concentration with photon subtraction on both modes.
      Contour plot (a) shows dependence of the logarithmic negativity on the local displacements $\alpha$, $\beta$,
      while in panel (b), the cut along $\beta = -\alpha$ is plotted for positive values of $\alpha$.
      The parameters used are $\lambda = 0.2$, $R_S = 0.1$, $\eta = 1$.
    }
  \end{figure}

  \begin{figure}
    \includegraphics[width=\linewidth]{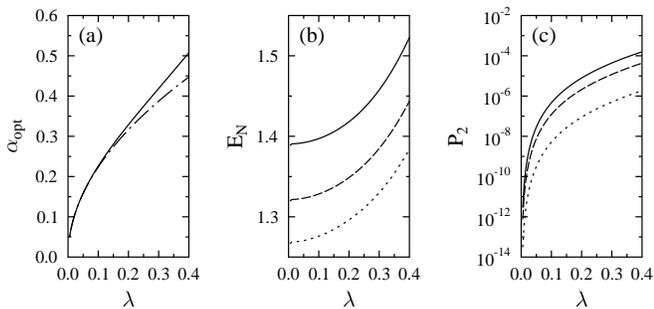}
    \caption{\label{fig.twomode_optimal}
      Optimal two-mode filtration in dependence on the squeezing constant $\lambda$.
      Panel (a) shows the optimal displacement $\alpha_\mathrm{opt}$, panel (b) represents the logarithmic negativity $E_N$,
      and in panel (c), we plot the success probability $P_2$.
      The results were obtained for $R_S = 0.1$, and $\eta = 1$ (solid line), $\eta = 0.5$ (dashed), $\eta = 0.1$ (dotted).
      The optimal displacement is mostly unaffected by limited detector efficiency;
      for better readability, only line for $\eta = 1$ is shown in panel (a).
      The dot-dashed line in this panel represents the $\sqrt{\lambda/2}$ reference curve,
      the optimal displacement for weakly squeezed input states.
    }
  \end{figure}

  The performance of entanglement concentration with photon subtraction on both modes is plotted in Fig.~\ref{fig.twomode}.
  The resulting dependence of entanglement on the coherent displacements $\alpha$ and $\beta$ 
  is similar to that of weakly squeezed states (cf. Fig.~\ref{fig.weak_two}).
  In what follows, we restrict ourselves to the diagonal line $\beta = -\alpha$ [Fig.~\ref{fig.twomode}(b)],
  at which the maximum value of logarithmic negativity lies. 
  Numerical simulations also indicate that the restriction to real displacements is not limiting,
  as complex phase factors do not increase the entanglement.  
  In Fig.~\ref{fig.twomode_optimal}, we investigate optimal entanglement concentration for different values of initial squeezing $\lambda$.
  The figure shows optimal displacement $\alpha_{\mathrm{opt}}$, the resulting logarithmic negativity $E_N$ and the success probability $P_2$.
  Similarly to the single-mode subtraction, the initial squeezing influences mainly the success probability
  while the logarithmic negativity only weakly increases with $\lambda$.
  Compared to the single-mode subtraction scheme, the amount of entanglement we can get from the system is higher,
  at a cost of smaller success probability. 
  Inefficient detection lessens the entanglement of the output state more than in the case of the single-mode subtraction,
  as two detectors are used. 
  As shown in Fig.~\ref{fig.twomode_optimal}(a), for $\lambda \lesssim 0.2 $ the optimal displacement $\alpha_{\mathrm{opt}}$ 
  practically coincides with the optimal value $\sqrt{\lambda/2}$ 
  determined analytically in the weak-squeezing limit, while for larger $\lambda$ the optimal displacement exceeds this analytical prediction.
  The success probability is the parameter that is affected most by imperfect detection and approximately scales as $P_{2}\propto \eta^2$.
  Obviously, this is the main disadvantage of the two-mode subtraction compared to the single-mode subtraction.

  \section{Conclusions}\label{sec.conclusions}

  In summary, we have analyzed continuous variable entanglement concentration based on photon subtraction enhanced by local displacements.
  We considered split single-mode squeezed vacuum state as input of the protocol, which makes the whole setup easier to implement experimentally
  and directly corresponds to the experiment by Takahashi \textit{et al.} \cite{Takahashi10}.
  We have investigated and compared two different scenarios --
  in the first one, the displacements and photon subtraction are performed on one mode only, which makes it experimentally easier,
  while the other scheme employs the filtration operation on both modes and thus introduces another degree of freedom
  that can be used to extract more entanglement in case of successful photon subtraction.

  In the case of the single-mode photon subtraction local displacements do not provide any improvement.
  From this point of view, simple photon subtraction, as already demonstrated by Takahashi \textit{et al.} \cite{Takahashi10}, is the best option,
  although other types of local operations, such as squeezing, might work differently.
  On the other hand, the scheme with photon subtraction on both modes can be significantly improved by local displacements.
  Moreover, the entanglement of the resulting state can be higher than in the case of the single-mode subtraction.
  This is a remarkable improvement, since single-photon subtraction can outperform photon subtraction on both modes without displacements.
  
  The required combination of photon subtraction and coherent displacement has already been successfully demonstrated experimentally
  and explored for engineering of arbitrary qubit states formed by superpositions of a squeezed vacuum and squeezed single-photon states 
  \cite{Nielsen10}.
  In this context, we note that the entanglement concentration scheme studied in this paper not only enhances the entanglement of the state but
  also modifies its structure. 
  Thus, the considered protocol can also be seen as an example of a bipartite quantum state engineering by local operations.
  In the future, it would be interesting to investigate the power and limitations of this approach and determine which states can be prepared 
  from initial Gaussian entangled states by local photon subtractions, local Gaussian operations and classical communication.

  \begin{acknowledgments}
    This work was supported by the Czech Science Foundation under project P205/12/0577
    and Grant No. PrF-2012-019 of the Palack\'y University.
  \end{acknowledgments}

  \bibliography{singlemode}
  
\end{document}